# Recurrence Quantification Analysis of Financial Market Crashes and Crises

Oleksandr Piskun[1,2], Sergii Piskun[3*]

[1]Academy of Fire Safety named after Chornobyl Heroes, Cherkasy, Ukraine
[2]Cherkasy National University named after B. Khmelnytsky, Cherkasy, Ukraine
[3]Cherkasy Banking Institute of Banking University of the National Bank of Ukraine, Cherkasy, Ukraine

Financial markets are systems with the complex behavior, that can be hardly analyzed by means of linear methods. Recurrence Quantification Analysis (RQA) is a nonlinear methodology, which is able to work with the nonstationary and short data series. Thus, we apply RQA for the studying of the critical events on financial markets. For the present research, stock crashes of DJI 1929; DJI, NYSE and S&P500 1987; NASDAQ 2000; HSI 1994, 1997 and Spanish 1992, Portuguese 1992, British 1992, German 1992, Italian 1992, Mexican 1994, Brazilian 1999, Indonesian 1997, Thai 1997, Malaysian 1997, Philippine 1997, Russian 1998, Turkish 2001, Argentine 2002 currency devaluations were taken. The recent world financial crisis of 2007-2010 was considered as well. The possibility of LAM measure to serve as a tool for the revealing, monitoring, analysing and precursoring of financial bubbles, crises and crashes was asserted.

## 1. Introduction

Modern economic system is presented by the real economy and financial sector. Financial sector is considered to serve the real economy. With the financial institutions development and capital flow acceleration, the possibility of earning money through speculative dealing at stock exchange occurred. When asset prices are essentially overrated, financial sector can loose contact with the real economy. Thereby, the bubble develops and leads to the crash in a sequel. In-turn it ruins not only speculators and investors, but also puts out of action the effective process of the real economy serving. Thus, crash can provoke a long-term recession.

During the last two decades the amount of crashes has increased in times. In addition, with globalization, financial crash of one country inevitably pulls collapses of numerous other countries. Therefore, an urgent need in the adequate tool of analysis and forecasting of such events on financial markets occurred.

Financial markets are fractal systems, which demand non linear methods for their description [Mandelbrot and Hudson, 2004]. One of these methods is Recurrence Plot (RP). It was introduced by Eckmann *et al.* [1987] in order to represent a system behavior on the basis of recurrence theory. Recurrence plot can be mathematically expressed as

$$R_{i,j}^{m,\varepsilon_i} = \Theta(\varepsilon_i - \left\| \vec{x}_i - \vec{x}_j \right\|), \vec{x} \in \Re^m, i,j = 1,...,N \qquad (1)$$

where $N$ is the number of considered states $x_i$, $\varepsilon_i$ is a threshold distance, $\|\cdot\|$ a norm and $\Theta(\cdot)$ the Heaviside function.

---

[*] *Corresponding author.* E-mail address: sergii_com@ukr.net

Zbilut and Webber [1992] developed a set of quantitative measures, that are calculated in the terms of RP. They defined recurrence rate (RR), determinism (DET), divergence (DIV), entropy (ENTR) and trend (TREND). These measures formed RQA [Trulla *et al.*, 1996]. Later, Gao [1999] proposed the time statistics measures: Trapping time 1 (T1) and Trapping time 2 (T2). Marwan [2002] worked out the measures of laminarity (LAM) and trapping time (TT).

RP and RQA are widely used in biology and physiology, but there are already works with applying to financial markets. Holyst *et al.* [2001], Gilmore [1992], McKenzie [2001] used these techniques for testing chaos in financial time series. Antoniou and Vorlow [2000] applied them for the indication of deterministic nonlinearities. Strozzi *et al.* [2002, 2007] used RQA for measurement of the financial data volatility and detection of correlation between currency time series. Fabretti and Ausloos [2005] applied RP and RQA for the critical regime detection in financial markets and estimation of the bubble initial time.

The aim of the present work is to reveal more abilities of RQA for the financial markets research, emphasizing on the RQA measures, quantified in dynamics. The study was conducted on various economic indexes, which contain crises or crashes.

The paper is ordered as follows. The second section presents the analysis of the historic stock crashes and recent world financial crisis 2007-2010. The study of the currency crises is shown in the third section. The fourth section contains the conclusion.

## 2. Stock market historic crashes and recent crisis analysis

Fabretti and Ausloos [2005] had already shown the ability of the visual and quantative recurrence analysis to distinguish the critical events on the financial markets from the normal regime. Let's see whether we can provide an effective financial bubbles analysis and forecasting by means of RQA. For the present study such stock exchange crashes as Dow Jones Industrial Average (DJI) 1929; DJI, NYSE and S&P500 1987; NASDAQ 2000; HSI 1994 and 1997 were taken (http://finance.yahoo.com/). All computations were done by means of MATLAB® version of RQA developed at the University of Potsdam called CRP toolbox, which can be found at http://tocsy.agnld.uni-postdam.de [Marwan *et al.*, 2007]

In [Piskun *et al.*, 2008] was shown that laminarity (LAM) is the most suitable measure, sensitive to critical events on markets. It is quantified as a ratio of quantity of the recurrence points in the vertical lines to all recurrence points in RP [Marwan *et al.*, 2007]. Actually, the inverse of laminarity reflects a market volatility level [Strozzi *et al.*, 2007].

First of all, let's have a look at the crash of DJI 1929 (see Fig. 1). The series of DJI quotes was taken in such a way, that the crash is in the middle. Window size – 250, dimension and delay – 1, radius $\varepsilon$ – 0,1. Embedding parameters were chosen empirically, as the most suitable for the calculation of measures in dynamics for our purpose. They will be used for any calculations through the article.

The calculated LAM measure begins from the horizontal coordinate x = 250, because the window size (250 points) was taken into account. Thereby, every point of LAM exactly corresponds to the DJI points.

According to the fig. 1, the LAM measure quantitatively estimates the regime of the market normal functioning. The decline of measure indicates the market instability that is promoted by the good trade expectations and, as a result, by volatility increasing. Horizontal coordinate x = 1355 indicates the border of instability, going through which, the crash is inevitable. The bubble is formed and the crash occurrence is just a question of the time. This border we are able to name as a "critical point". LAM also shows the time of the full market relaxation. A period between the critical point and full market relaxation, that contains crash and economic depression, we define as a "critical period".

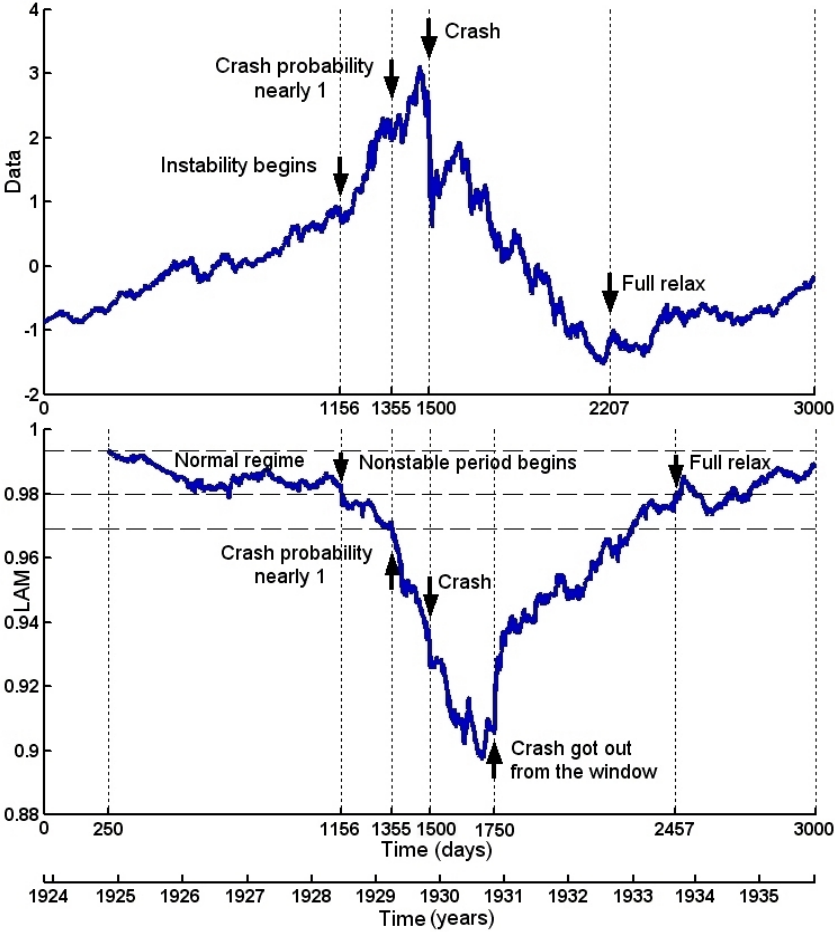

Fig. 1. Dow Jones Industrial Average rates 1924-1935 and its correspondent laminarity.

After 04.05.1928 (x = 1156) LAM began to fall down, that indicates the beginning of instability. The critical point appeared 26.03.1929 (x = 1355). Thus, LAM had given signals for 342 and 145 trading days before the crash (there are 250 trading days in the year). This time is more than enough for the acceptance of arrangements for elimination of the bubble burst.

The point of the full market relaxation occurred 18.08.1932 (x = 2207), because the segment of the stock series, engulfed by the window (x = 2207 - 2457), was already functioning in its normal regime. Therefore, the full relaxation started at the first point of the window.

It is necessary to show the quantitative description of the market conditions. But the different rows will give the different absolute measure values, therefore, it is better to conduct estimations in percent to the lowest border of the normal regime (see table 1).

Table 1. Quantitative characteristics of Dow Jones Industrial crash 1929.

|  | Normal regime | Instability begins | Critical point | Crash | Min | Full relaxation |
|---|---|---|---|---|---|---|
| Coordinate x | 0 - 1155 | 1156 | 1355 | 1500 | 1700 | 2207 |
| Days to crash | - | 342 | 145 | - | - | - |
| LAM | 0.980-0.992 | 0.980 | 0.969 | 0.925 | 0.898 | - |
| LAM, % | - | - | 1.122 | 5.612 | 8.367 | - |

For the extraction of a regularity of the LAM behavior, we need to provide estimations of other crashes. For this, we can't take such a long series as 3000 values, because they will contain, besides the analysed crash, other critical events. This can lead to the distortion of the estimation. So the length of the rows was chosen 1000 points. The crash is also not in the middle, but the row is cut in such a way to contain all regimes of functioning (see Fig. 2).

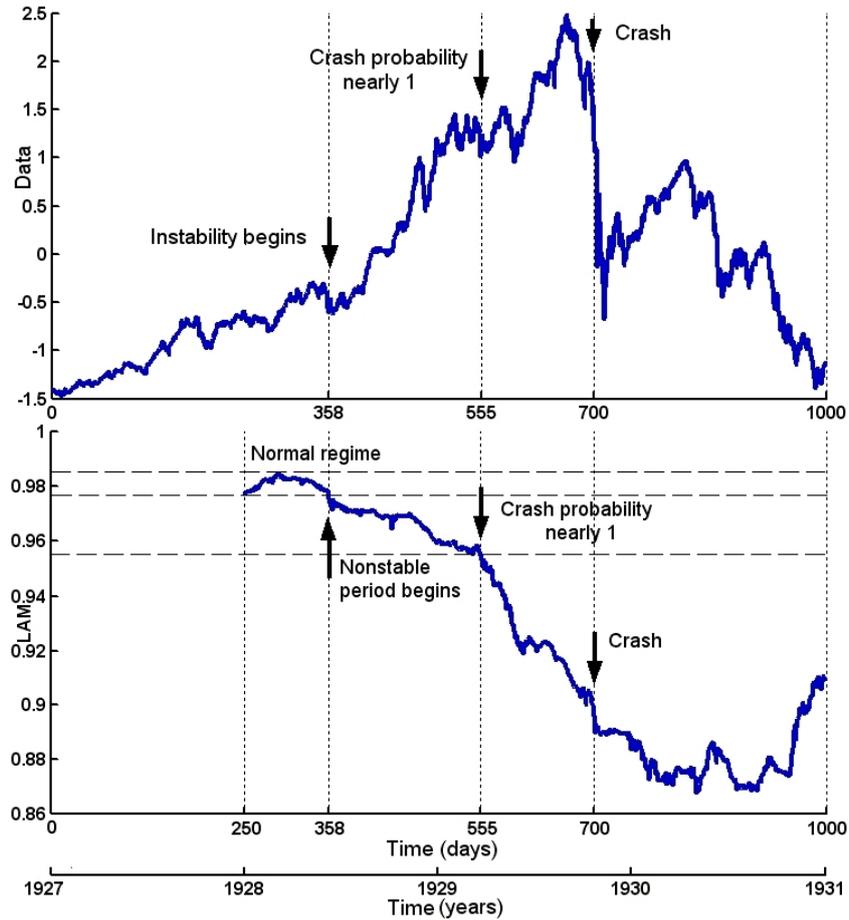

Fig. 2. Dow Jones Industrial Average rates 1927-1930 and its correspondent laminarity.

As we can see, the different regimes of functioning and the amount of days to the crash are not changed. This indicates the stability of the obtained results and one more time proves the noncriticality of RQA measures to the data length.

Let's have a look at the quantitative characteristics of the analysed crashes (see table 2).

Table 2. Quantitative characteristic of the analyzed crashes.

|  | Normal regime | Instability begins | Critical point | Crash |
|---|---|---|---|---|
| DJI 1929 | | | | |
| Days to crash | - | 443 | 246 | - |
| LAM | 0.985-0.977 | 0.977 | 0.955 | 0.872 |
| LAM, % | - | - | 2.25 | 10.747 |
| DJI 1987 | | | | |
| Days to crash | - | 341 | 161 | - |
| LAM | 0.989-0.985 | 0.985 | 0.964 | 0.932 |
| LAM, % | - | - | 2.178 | 5.380 |
| NYSE 1987 | | | | |
| Days to crash | - | 554 | 150 | - |
| LAM | 0.995-0.991 | 0.991 | 0.966 | 0.931 |
| LAM, % | - | - | 2.523 | 6.054 |
| S&P 1987 | | | | |
| Days to crash | - | 352 | 150 | - |
| LAM | 0.987-0.976 | 0.976 | 0.963 | 0.932 |
| LAM, % | - | - | 1.332 | 4.721 |
| HSI 1994 | | | | |
| Days to crash | - | 365 | 20 | - |
| LAM | 0.989-0.981 | 0.981 | 0.966 | 0.957 |
| LAM, % | - | - | 1.529 | 2.509 |
| HSI 1997 | | | | |
| Days to crash | - | 210 | 43 | - |
| LAM | 0.967-0.959 | 0.959 | 0.927 | 0.895 |
| LAM, % | - | - | 3.337 | 6.674 |
| NASDAQ 2000 | | | | |
| Days to crash | - | 343 | 11 | - |
| LAM | 0.989-0.977 | 0.977 | 0.937 | 0.925 |
| LAM, % | - | - | 4.094 | 5.322 |

According to the present table we can claim that:

- nonstable period begins much more earlier than the crash. In our case it began from 210 (HSI 1994) to 443 (DJI 1929) days before. This time is more than enough to provide the monetary regulations and blow out the bubble;

- the critical point occurs closer to the crash, from 246 in 1929 (DJI) to 11 in 2000 (NASDAQ) days before. It means the ability of quick capital consolidation and investment in a profitable market nowadays;

- LAM reveals different features of the market behavior. It shows, that DJI 1987 and NYSE 1987 crashes are similar, but S&P 1987 is rather weaker, because the drop of it's LAM is twice as high as the DJI's and NYSE's. Similarly, the crash of HSI 1997 is approximately in 2.5 times stronger than HIS 1994.

- LAM downfalls from the normal regime to the critical and crash points are different in different crashes, because the volatility of various markets in various times isn't the same. Thus, we can't predict the exact crash time by means of RQA, but it suits for the analysis of the financial markets.

After the analysis of the former crashes, let's apply RQA for the 2007-2010 crisis. Rows of DJI (27.07.2005-11.07.2011), S&P (27.07.2005-11.07.2011), FTSE (02.08.2005-11.07.2011), DAX (26.08.05-11.07.2011), HSI (26.07.05-11.07.2011), Nikkei 225 (21.05.05-11.07.2011), PFTS (11.08.06-11.07.2011), RTSI (21.06.05-11.07.2011) were taken. The length of each row is 1500 days. At Fig. 3 the DJI rates are presented.

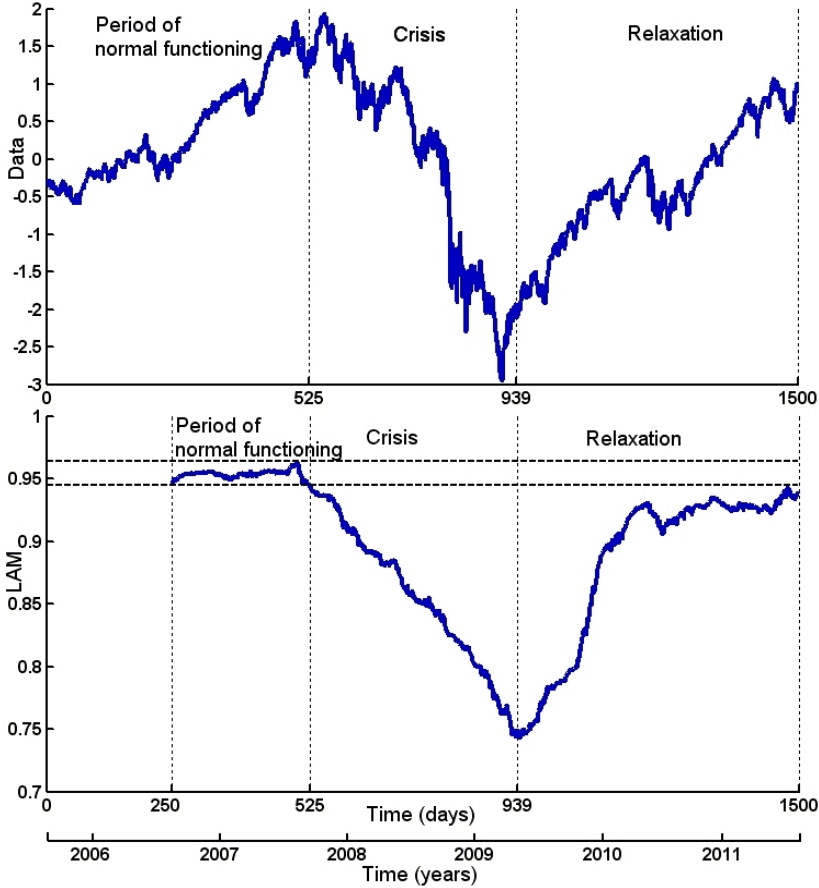

Fig. 3. Dow Jones Industrial rates 2005-2011 and its correspondent laminarity.

As we see, the view of the recent crisis is different from the former crashes. LAM also differs and doesn't contain a critical point. As such event as crash does not take place, the window of the LAM calculation has no rapid drop of the normalized data. So, LAM is able to detect the point of the relaxation beginning, because it has a smooth increasing without any gaps.

The whole world recognized the crisis existence after the Bear Stearns default in the middle of March 2008. In 2009 experts claimed that the crisis of banking liquidity had begun in the middle of summer 2007, but at that time nobody knew that. The measure signalized about the crisis beginning 28.08.2007 (x = 525). So, RQA is a suitable method for the adequate identification of crisis starting. The increasing of LAM after 23.04.2009 (x = 939), reflects the beginning of the market relaxation. The result of other indexes analysis is presented in table 3.

Table 3. Time characteristics of current crisis.

| Index | Crisis begins | Relaxation begins |
|-------|---------------|-------------------|
| DJI   | 28.08.2007    | 23.04.2009        |
| FTSE  | 07.08.2007    | 02.02.2009        |
| S&P   | 30.07.2007    | 18.05.2009        |
| DAX   | 19.11.2007    | 08.05.2009        |
| HSI   | 27.07.2007    | 09.01.2009        |
| Nikkei| 03.08.2007    | 27.01.2009        |
| PFTS  | 22.08.2007    | 15.09.2008        |
| RTSI  | 10.12.2007    | 22.12.2008        |

For DJI, FTSE, HSI and Nikkei, crisis started approximately simultaneously, as their business cycles and economics are highly synchronized. PFTS demonstrated the crisis events even before the DJI, as the capital running from emergent market occurred. RTSI and DAX began to feel the critical situation a little bit later, because of the lag in the economic cycles. Relaxation started in different countries at diverse times. Each government provided its own measures to eliminate the crisis, according to its policy and aims. Thus, the relaxation beginning can't be similar.

Speaking about the current situation on the financial markets, we can hardly claim about the full relaxation, because its point appears at the LAM graph only a year later after its factual existing (see fig.1). Moreover, after the crisis, markets can change their behaviour significantly. This way, a new level of LAM, that indicates the normal market functioning, could be set up. According to the present study, we can say, for sure, that DJI, S&P500, FTSE, N225 and HIS are rather close to their full relaxation or already gained it, some time earlier. We have a different situation with DAX (see fig. 4).

The period from 26.08.2005 till 16.11.2007 (x = 0 - 570) we can formally name the period of the normal functioning, because DAX has rather turbulent time before and did not fully recreated. The crisis lasted from 19.11.2007 till 07.05.2009 (x = 571 - 940). Then, DAX experienced relaxation from 08.05.2009 till 21.02.2011 (x = 941 - 1401). Because of the economic problems in Euro zone, DAX reflects the crisis events from 22.02.2011 till 11.07.2011 (x = 1402 - 1500). Under the "crisis events" we mean the increasing of market volatility, because of the negative trade expectations, and going out from the normal regime of functioning.

We have rather interesting situation with the Ukrainian PFTS index (see fig. 5).

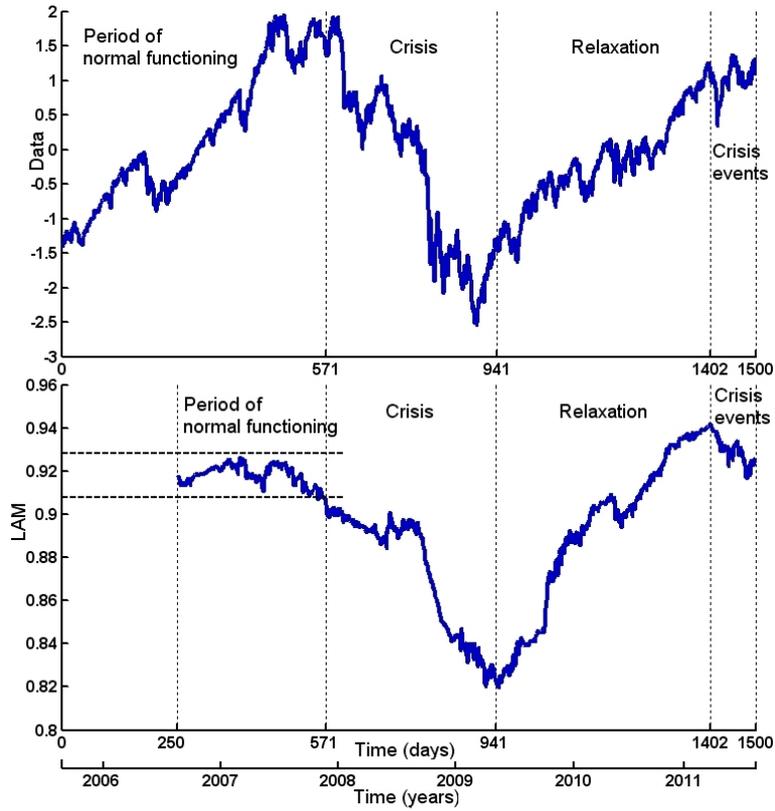

Fig. 4. DAX rates 2005-2011 and its correspondent laminarity.

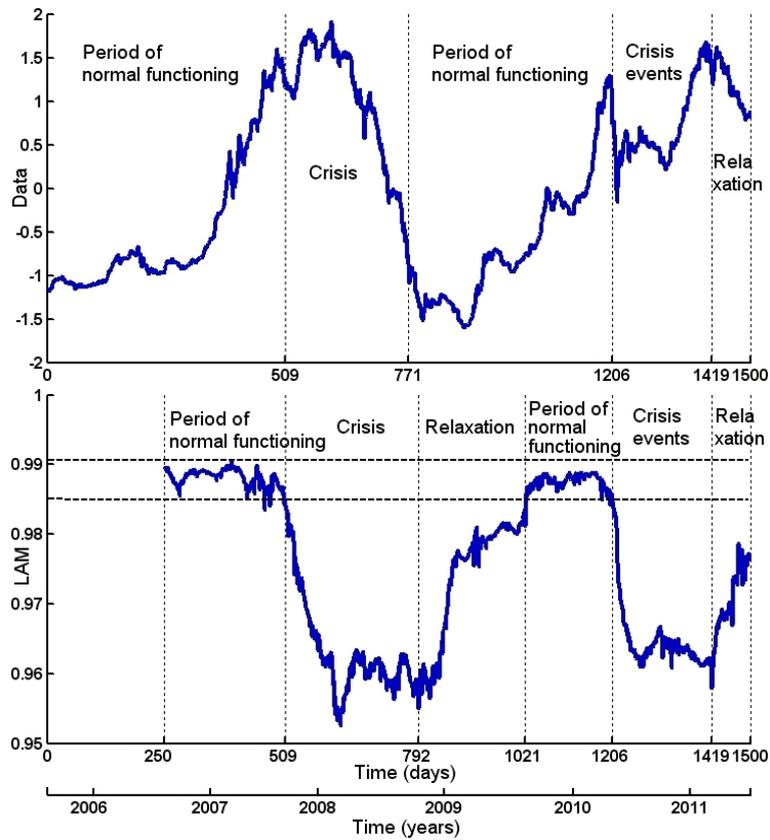

Fig. 5. PFTS rates 2005-2011 and its correspondent laminarity.

For PFTS the crisis began 22.08.2007 (coordinate x = 509). According to the lower graph (LAM), the point of the full relaxation is x = 1021. Taking into account the window size – 250, a real point of the full relaxation is x = 771 at the upper graph. Its even earlier than the relaxation period began (x = 792). In such case, the relaxation did not exist. Because of specific of the Ukrainian emergent stock market, that does not reflect the sector of the real economy, when the crisis ended – it just moved to the normal functioning. As we have a window-based system of calculation, the window need some time to be cleared up from the crisis points that make the LAM lower. That is why, we can claim, that 15.09.2008 (x = 771) the crisis ended and the market shifted to the normal regime of functioning. 07.05.2010 (x = 1206) crisis events took place, because of the unstable politic situation in the country and the profit fixation of portfolio investors. From 14.03.2001 (x = 1419) the relaxation of the market can be observed.

RTSI is also not a standard case for the analysis (see fig. 6).

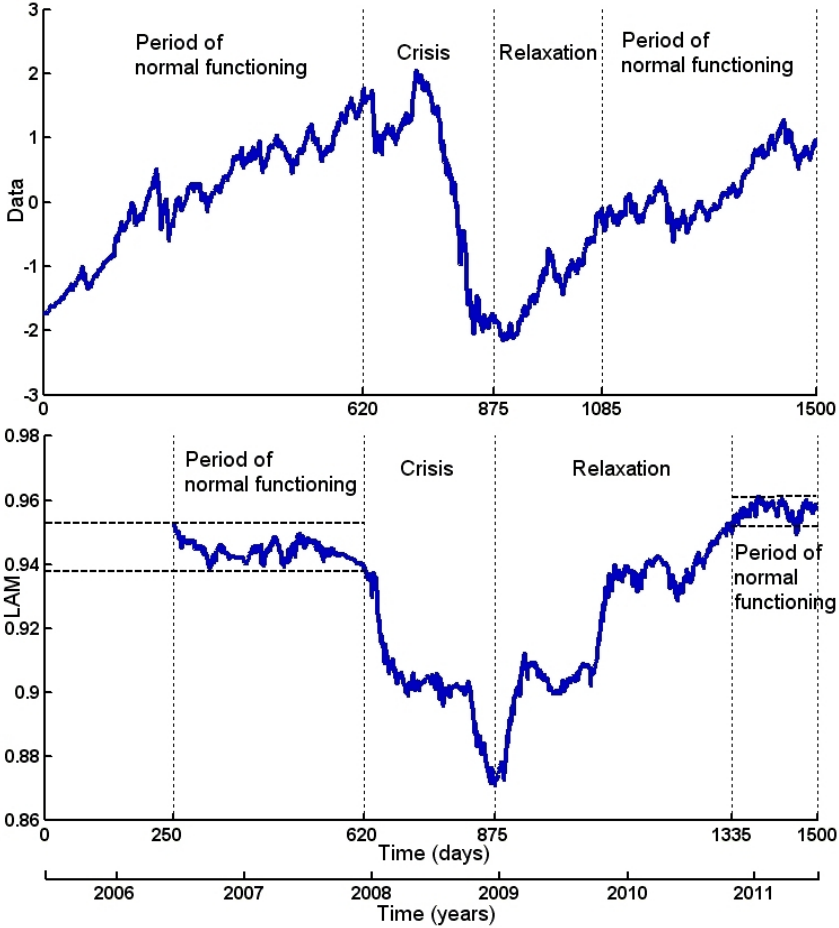

Fig. 6. RTSI rates 2005-2011 and its correspondent laminarity.

The Russian index experienced crisis from 10.12.2007 (x = 620) till 19.12.2008 (x = 874) and relaxation from 22.12.2008 (x = 875) till 26.10.2009 (x = 1084). A new regime of the normal functioning (x = 1335 – 1500 at lower graph) characterized by the higher LAM values, than the previous one (x = 250 – 619 at lower graph), because the market changed its behavior after the crisis and became less turbulent. Moreover, RTSI is the first index that fully relaxed, 27.10.2009 (x = 1085).

So, according to the present study, RQA is a good instrument for the stock market analysis. Therefore, we can assume that it is useful for monitoring and forecasting. First, consider the regular procedure of RQA calculation. In the standard method, the row of date is taken, then the RP is built and RQA measures are calculated. This way, every point of RQA measure, except the last one, is calculated with the involving of the future points of the row. For example, there are DJI quotes from 09.08.1999 till 11.07.2011 (see fig. 7.). The 1375-th point of LAM (x = 1625) is calculated on the basis of the past and the future points of DJI. The last point of LAM (x = 3000) uses only the past points of DJI.

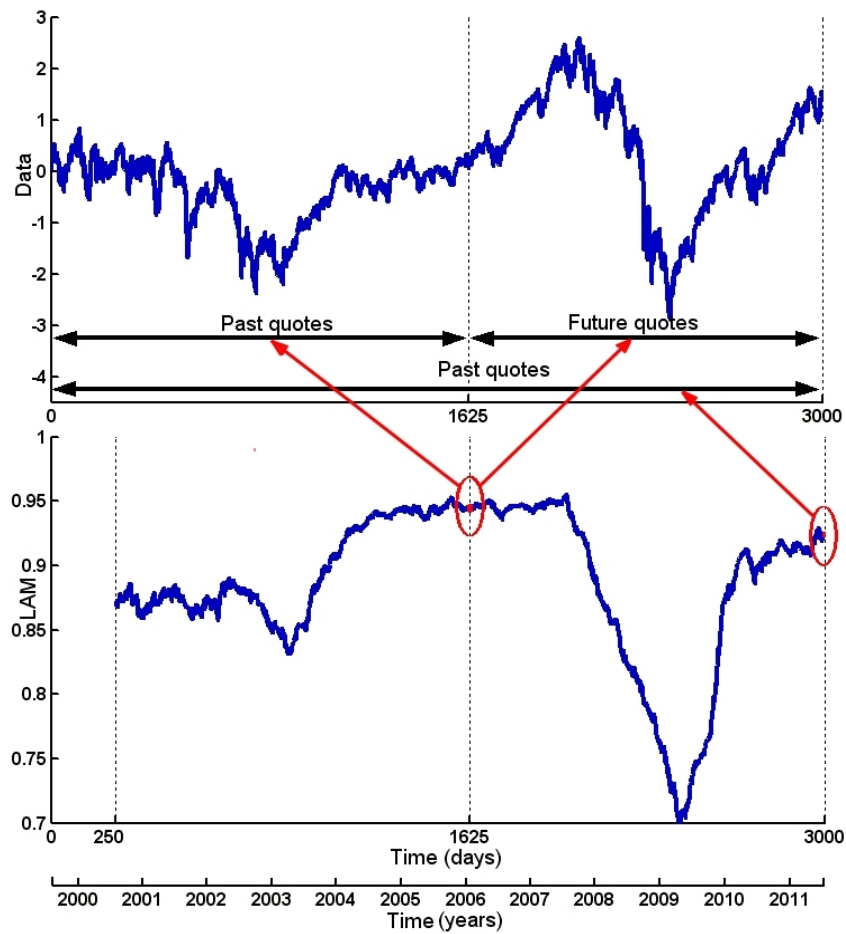

Fig. 7. DJI rates 1999-2011 and its correspondent laminarity.

Thus, in order to simulate the process of everyday monitoring, we changed the method of the RQA calculation. First of all, we take a row of the financial data, as usual. Then, the part of the row is cut off. For this part, we build a recurrence plot and provide the calculation of LAM. But only the last point of the LAM graph is taken and written down. After this, we choose the next part of the row with the step of 1 day, cut it off and repeat the procedure. This way, we obtain the LAM graph like in the case of everyday calculation in real time. If such LAM reveals the same regimes of market functioning as the regular one, we will be able to claim, that RQA is suitable for the monitoring of the market tendencies.

A new method of RQA calculation brings in an additional parameter – the length of the part of the row, which is cut off (LPR). Fig. 8 shows LAM graphs calculated with different LPR in comparison with the regular LAM.

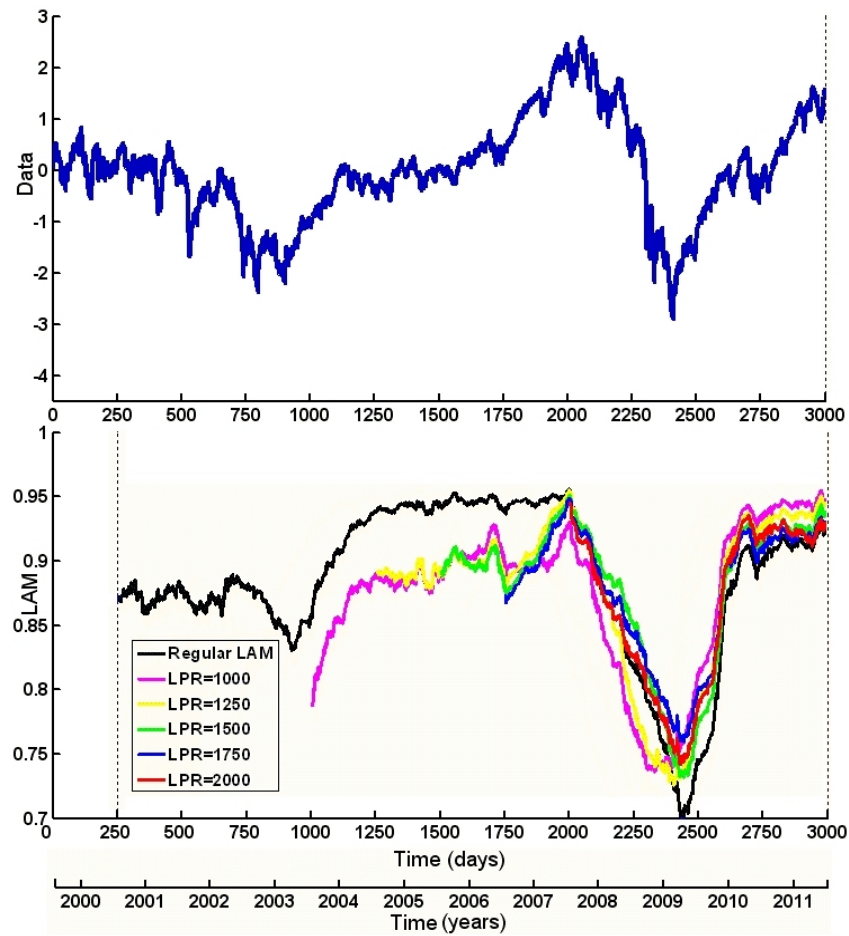

Fig. 8. DJI rates 2005-2011 and its various LAMs.

When LPR is 1500, 1750 or 2000, the new LAMs are rather similar to the regular one. So, the minimum LPR should be 1500 points while embedding dimension and delay = 1, threshold = 0,1, window size = 250.

The new LAM with LPR = 1500 gives the same points of crisis and relaxation beginning as the regular one (see fig. 9).

The new LAM (red one), can't define the period of the normal functioning like a regular LAM (x = 1500 - 2000), because LPR engulfs the former crisis of 2000. Though, providing monitoring in real time, we are able to identify the crisis 28.08.2007 (x = 2026) and relaxation 23.04.2009 (x = 2441) beginning. The part of LAM x = 2441 - 3000 gives us the information about the active recreation of the market. These results satisfy the aim of the effective monitoring.

Forecasting of the crisis and recreation is limited. As every crisis or crash is different (in terms of volatility) according to the table 2, and the new LAM does not reveal any additional information, we can only assume the high probability of such events some time in future in respect to the LAM behavior, but are not able to predict the exact dates.

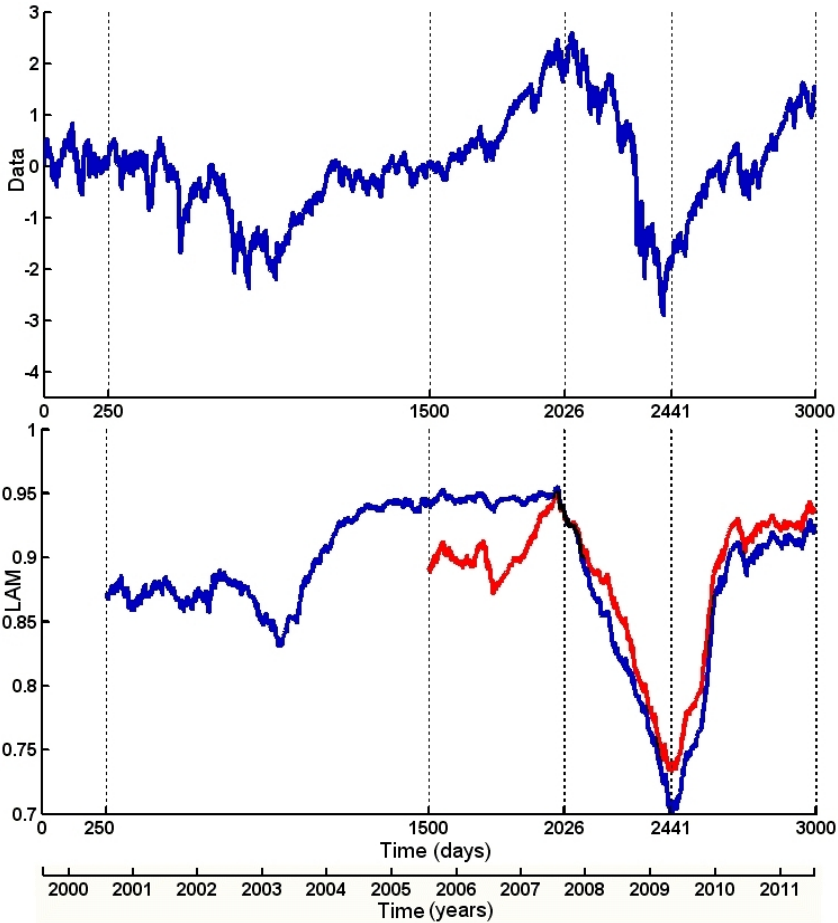

Fig. 9. DJI rates 1999-2011, its correspondent regular and new laminarities.

## 3. The analysis of the currency markets crises

After the stock markets study, let's see whether RQA is suitable for the currency crisis analysis. For this, the Spanish 1992, Portuguese 1992, British 1992, German 1992, Italian 1992, Mexican 1994, Brazilian 1999, Indonesian 1997, Thai 1997, Malaysian 1997, Philippine 1997, Russian 1998, Turkish 2001, Argentine 2002 crises were taken (http://www.oanda.com/currency/historical-rates).

According to the provided research, two types of currency crises were revealed. The first one occurs when the central bank carries out a tough monetary policy, and because of the overestimated currency exchange rate, economic troubles and speculative attacks, it can't support the high rate any more. In such case the devaluation takes place. The currency rate becomes flexible and after some crisis period moves to the normal free floating regime of functioning.

All analysed crises, besides the European one 1992, refer to the first type. For example, let's have a look at the Indonesian 1997 and Brazilian 1999 crises (see Fig. 10, 11).

We consider that crisis contains two phases: economic decreasing and recreation. So, we include the period of relaxation into the period of the crisis. The Indonesian crisis began 23.08.1997 (x = 995), but LAM measure could not signalize about it some time before, because of the tough currency rate policy. The devaluation was 05.01.1998 (x = 1130); 13.08.1998 (x = 1350) the relaxation began and 05.01.1999 (x = 1495) market moved to the normal free floating regime.

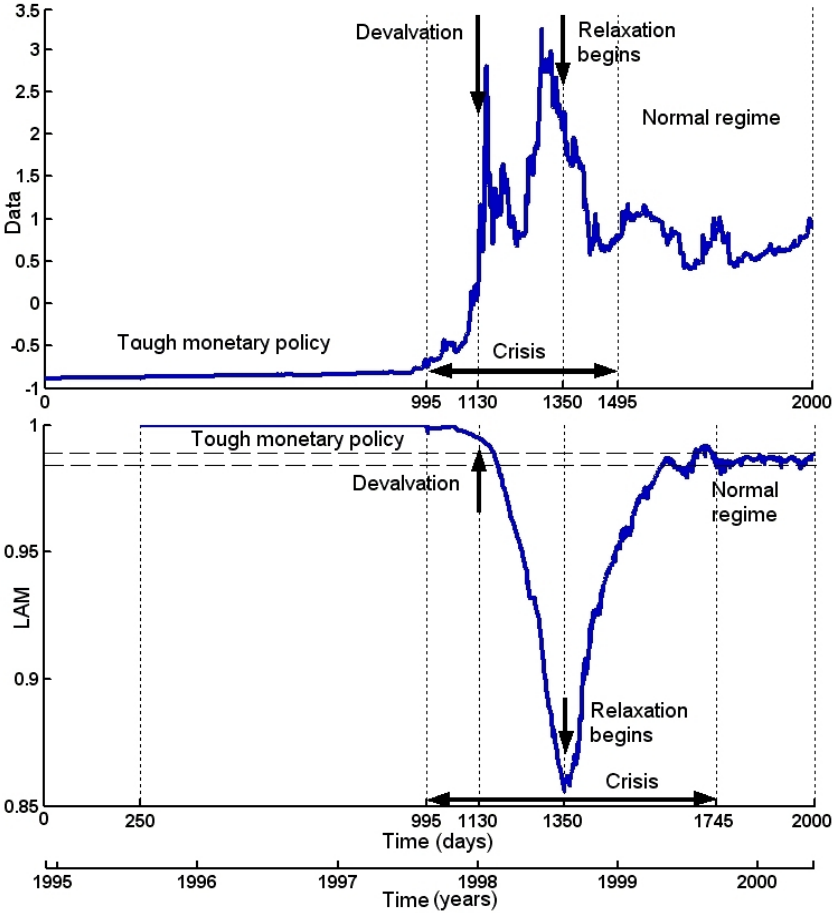

Fig. 10. Indonesian rupiah rates 1995-2000 and its correspondent laminarity.

The Brasilian crisis began with devaluation 04.01.1999 (x = 1008). The beginning of relaxation 14.09.1999 (x = 1250 at the lower graph) coincided with the end of the crisis (x = 1500 at the lower graph, but x = 1250 at the upper one). It means that market didn't need the period of crisis damping and just moved to the normal regime of functioning. Unfortunately, we can state this fact only post facto and are not able to forecast it somehow. The LAM measure falls 05.08.2000 (x = 1837), signalling about the Brasilian crisis 2001.

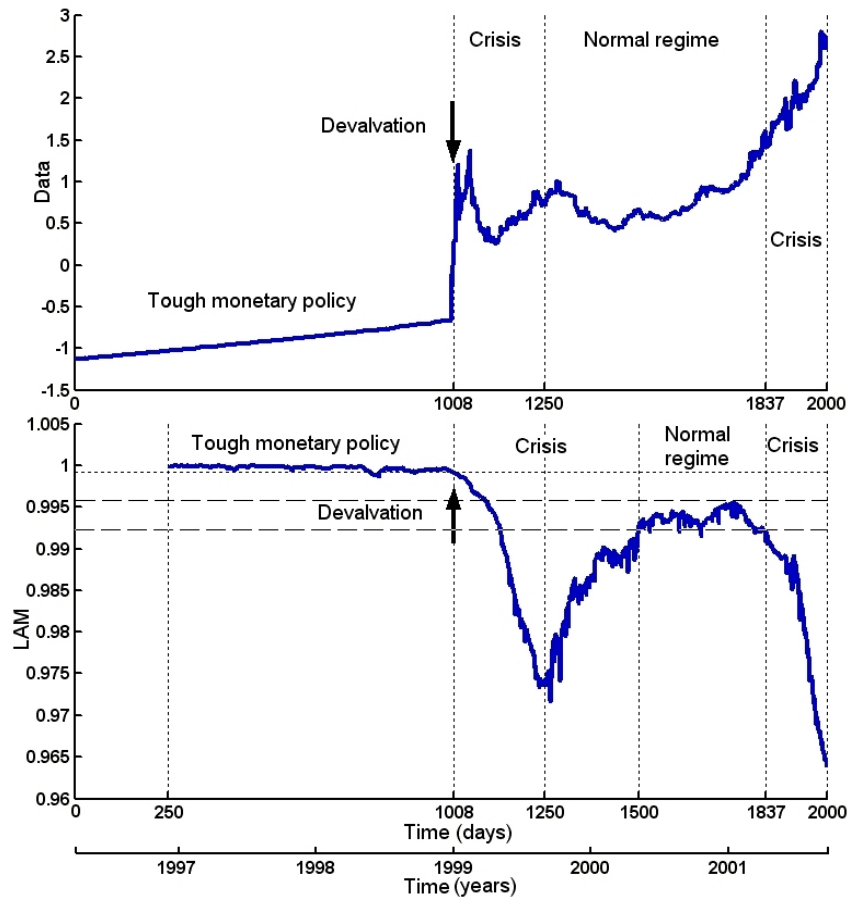

Fig.11. Brazilian real rates 1996-2001 and its correspondent laminarity.

Thus, RQA is able to detect and analyse crises of the first type, but can't predict them, because of the tough or even fixed regime of currency rates.

The second type of the currency crisis is when the central bank provides free floating regime and sometimes supports it with the intervention or interest rate changing, but because of the rate overestimation and speculative attacks, currency can loose a significant portion of its value in short time. The European crisis 1992 refers to this type. Let's have a look at the British pound rates from 11.09.1989 till 03.03.1995 (see Fig. 12).

For the British pound, crisis began 04.12.1990 (x = 450). The devalvation occurred 14.09.1992 (x = 1100). The relaxation began 11.05.1993 (x = 1339) and 29.09.1993 (x = 1480) the market moved to the normal regime. The view of LAM measure is different from the LAM of the stock crash, thus we can't identify the point of devalvation, for example like the critical point. Though, laminarity is still able to reveal the moments of the crisis and relaxation beginning, crisis ending. The results of analysis of other currencies are presented in table 4.

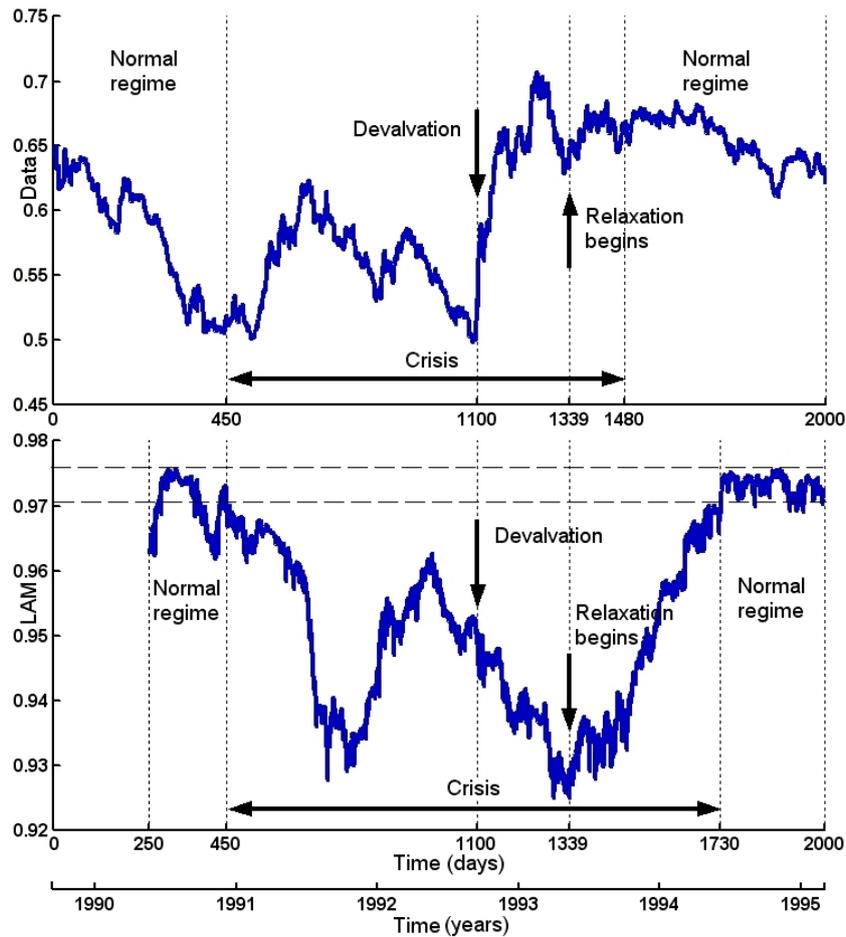

Fig. 12. British pound rates 1989-1995 and its correspondent laminarity.

Table 4. Time characteristics of the analyzed currency crises.

|  | Crisis begins | Devalvation | Relaxation begins | Crisis end (full relaxation) |
|---|---|---|---|---|
| Great Britain | 04.12.1990 | 14.09.1992 | 11.05.1993 | 29.09.1993 |
| Germany | 20.12.1990 | - | 19.02.1993 | 04.10.1993 |
| Italy | 08.03.1991 | 13.09.1992 | 13.05.1993 | 04.07.1993 |
| Finland | 05.06.1991 | 14.11.1991 | 10.05.1993 | 21.08.1993 |
| Swedish | 04.05.1991 | 20.11.1992 | 12.05.1993 | 10.05.1995 |
| Portugal | 17.12.1990 | 14.09.1992 | - | 18.08.1994 |
| Spain | 08.10.1990 | 14.09.1992 | - | 25.10.1993 |

In Great Britain, Germany, Portugal and Spain crisis began in 1990. Later, in 1991 it began in Italy, Finland and Sweden. The first devalvation was in Finland 1991, other countries experienced it in 1992. In the case of Germany, it had only timely falling of the currency rate. The relaxation of the markets began in 1993, approximately at the same time.

Exceptions are Portugal (fig. 13) and Spain. The currency markets gained the full relaxation at the end of 1993. Only Sweden and Portugal got out of the crisis later.

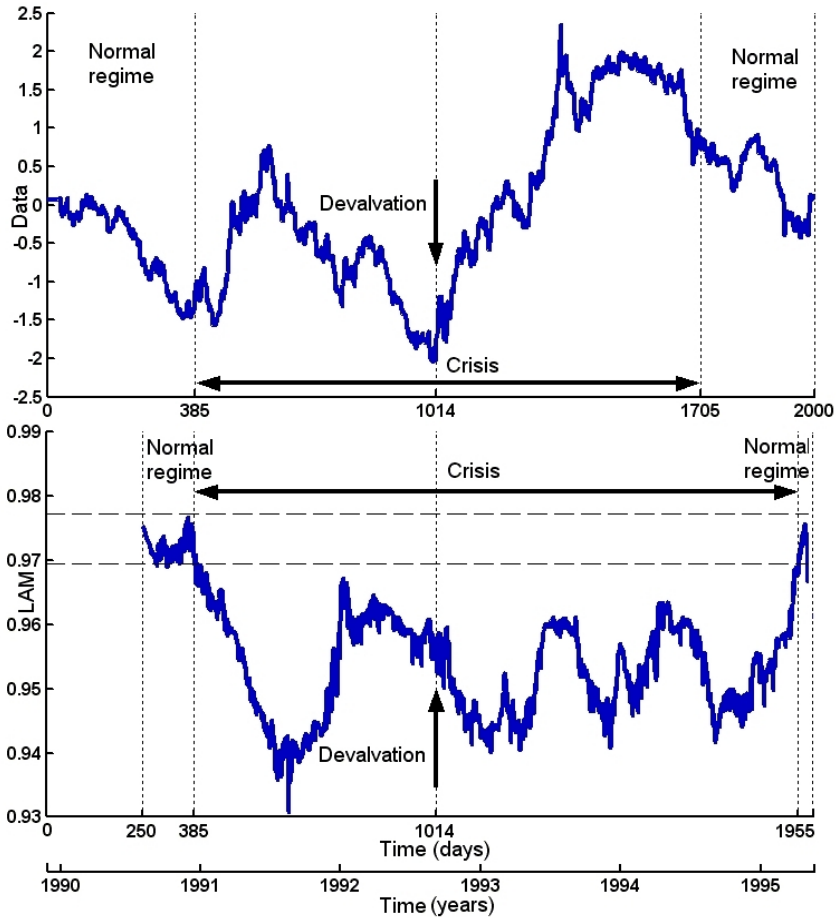

Fig. 13. Portuguese escudo rates 1989-1995 and its correspondent laminarity.

Fig. 13 shows that we can't exactly define the point of relaxation beginning, because of the constant market turbulence and strong volatility jumps. The same situation we have with the Spanish peseta.

After the study of the former crises, let's apply RQA for the detection of the crisis 2007-2010 influence on the currency market. For this, the EUR/USD pair from 20.09.2006 till 01.07.2011 was taken (see Fig. 14).

The crisis for EUR/USD began 20.01.2008 (x = 415). It's 4 monthes later than for DJI (28.08.2007). 24.04.2009 (x = 810) the relaxation started. It's only a day later than at DJI (23.04.2009). From 29.04.2010 (x = 1125) the crisis events are observed, that might be connected with the problems in Euro zone. So, the crisis has an influence on the currency market with some delay in comparison with the stock market.

In order to check the possibility of monitoring and forecasting of the currency market behaviour, we applied the same procedure, as in chapter 2 for the stock market, to the EUR/USD pair from 19.12.2002 till 11.07.2011.

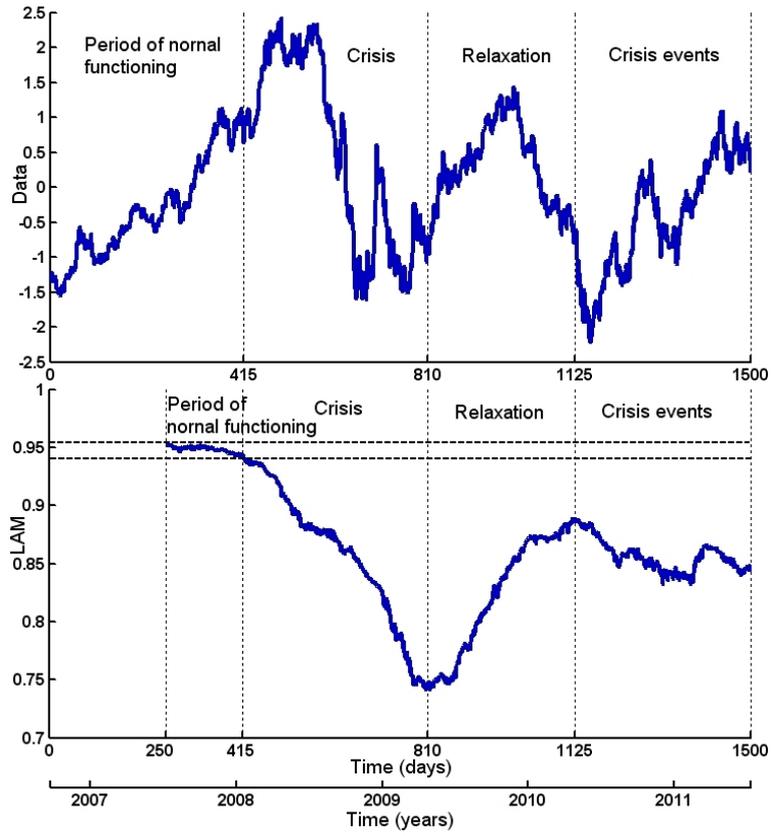

Fig. 14. Currency pair EUR/USD rates 2006-2011 and its correspondent laminarity.

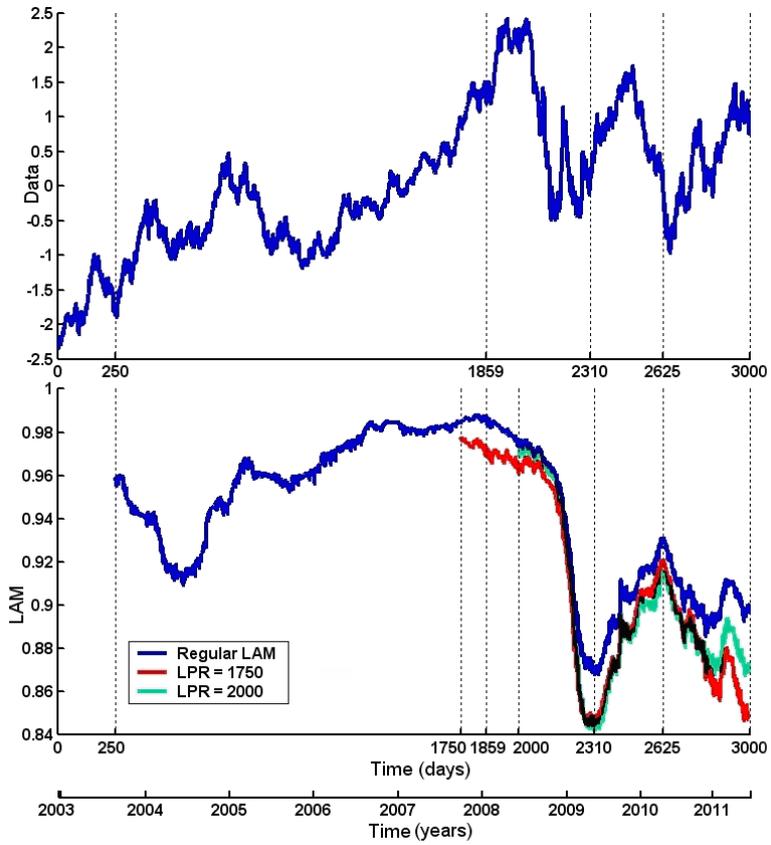

Fig. 15. EUR/USD rates 2002-2011, its correspondent regular and new laminarities.

After the choosing of different LPR, we came to the conclusion that LPR = 1750 is a minimum length for the currency market (see fig. 15). Because of the high turbulence of such kind of the financial markets, we should take longer LPR than in the case of stocks.

Because of rather long LPR, we can't estimate the point of crisis starting (20.01.2008, x = 1859), but we are able to catch further occurrences – the relaxation (24.04.2009, x = 2310) and crisis events (29.04.2010, x = 2625) beginning. So, by means of the new way of LAM calculation, we can provide monitoring of the currency markets. The process of forecasting is also limited similarly to the case of the stock markets.

## 4. Conclusion

The analysis of historic crashes of the stock markets revealed the possibility of RQA, especially LAM measure, to distinguish diverse market periods: normal functioning, instability, critical period, relaxation. As LAM is rather sensitive to the market state changes, detection of instability and critical point in time can signalize about the negative tendencies long before the crash occurrence. A common pattern of various crashes, based on the volatility was not detected. Thus, it is impossible to predict the crash date by means of LAM.

The study of stock markets during the 2006-2010 crisis showed that markets have different periods of functioning than in the case of the crash. They are: the period of normal functioning, crisis, relaxation and existing of the crisis events. The current situation in the stock markets is: DJI, S&P500, FTSE, N225 and HIS approximately gained the point of the full relaxation or already functioning in the normal regime; DAX experiences crisis events; RTSI is fully relaxed and PFTS is recreating.

The exploration of the historic currency crises exposed two types of them. The first one concerns the market shifting from the fixed regime of functioning to the free floating one. The second type reflects the currency crisis, during the free floating regime, that occurs because of the economic problems and trade speculative attacks. The eur/usd analysis showed that the 2007-2010 crisis had an influence on the currency market with some delay in comparison to the stock one. Speaking about the current situation, eur/usd experiences the crisis events.

In order to check the RQA possibility of the financial market monitoring, we proposed a different procedure of LAM calculation. The difference is that each point of the new LAM requires only the former data points. As the new LAM has rather similar view as the regular one, it can be used for the monitoring of stock and currency markets in real time.

Acknowledgment: the authors are grateful for Dmitry Chabanenko, Cherkasy National University named after B. Khmelnytsky, Cherkasy, Ukraine, for the development of the toolbox with the new procedure of RQA measures calculation.